\documentclass[printer]{aa}
\usepackage{lineno}
\usepackage{natbib}
\bibpunct{(}{)}{;}{a}{}{,} 
\usepackage{graphicx}
\usepackage{txfonts}
   \usepackage{color}  

\begin{document}

\title{Long-Term Multiwavelength Studies of High-Redshift Blazar 0836+710}

 \author{A. Akyuz
          \inst{1, 2}
          \and
          D. J. Thompson\inst{2}
          \and
         D. Donato\inst{2, 3}
         \and
         J. S. Perkins\inst{2, 4}
         \and
         L. Fuhrmann\inst{5} 
         \and
         E. Angelakis\inst{5}
         \and
         J. A. Zensus\inst{5}
         \and
         S. Larsson\inst{6, 7, 8 }
          \and
          K. Sokolovsky\inst{9, 10}
         \and
         O. Kurtanidze\inst{11, 12, 13}}

   \institute{University of Cukurova, Department of Physics, 01330 Adana, Turkey\\
              \email{aakyuz@cu.edu.tr}
         \and
             NASA Goddard Space Flight Center, Greenbelt, MD 20771, USA\\
              \email{David.J.Thompson@nasa.gov}
           \and
             CRESST, Department of Astronomy, University of Maryland, College Park, MD 20742, USA\\
              \and
              CRESST, CSST, University of Maryland, Baltimore County, Baltimore, MD 21250, USA\\
              \and
              Max-Planck-Institut f\"ur Radioastronomie, Auf dem H\"ugel 69 53121 Bonn, Germany\\
              \and
              Department of Physics, Stockholm University, AlbaNova, SE-106 91 Stockholm, Sweden\\
               \and
               The Oskar Klein Centre for Cosmoparticle Physics, AlbaNova, SE-106 91 Stockholm,
Sweden\\
\and
Department of Astronomy, Stockholm University, SE-106 91 Stockholm, Sweden\\
               \and
               Astro Space Center of Lebedev Physical Institute, Profsoyuznaya Str. 84/32, 117997 Moscow, Russia\\
               \and
       Sternberg Astronomical Institute, Moscow State University, Universitetskii~pr. 13, 119992 Moscow, Russia\\
              \and       
             Abastumani Observatory, 383762 Abastumani, Republic of Georgia.\\  
             \and
              Landessternwarte, Zentrum f\"ur Astronomie der Universit\"at Heidelberg, K\"onigstuhl 12, 69117 Heidelberg, Germany\\
              \and
		Engelhardt Astronomical Observatory, Kazan Federal University, Tatarstan, Russia
 }

\date{     }
  
\abstract
   {}
   {The observation of $\gamma$-ray flares from blazar 0836+710 in 2011, following a period of quiescence, offered an opportunity to study correlated activity at different wavelengths for a high-redshift (z=2.218) active galactic nucleus.}
   {Optical and radio monitoring, plus {{\it Fermi}}-LAT $\gamma$-ray monitoring provided 2008-2012 coverage, while {{\it Swift}} offered auxiliary optical, ultraviolet, and X-ray information. Other contemporaneous observations were used to construct a broad-band spectral energy distribution. }
   {There  is evidence of correlation but not a measurable lag between the optical and $\gamma$-ray flaring emission.  On the contrary, there is no clear correlation between radio and $\gamma$-ray activity, indicating radio emission regions that are unrelated to the parts of the jet that produce the $\gamma$ rays.  The $\gamma$-ray energy spectrum is unusual in showing a change of shape from a power law to a curved spectrum when going from the quiescent state to the active state.  }
   {}

\keywords{galaxies:active-gamma rays: galaxies-quasars:individual (0836+710)}

\maketitle


\section{Introduction}
The luminous high-redshift \citep[$z$=2.218;][]{Stickel1993} quasar of the blazar sub-class 0836+710 (also known as S5 0836+71 or 4C +71.07)
 is characterized by a flat radio spectrum \citep[$\alpha$=$-$0.33;][]{Kuhr81}. It hosts a powerful one-sided radio jet emerging from the core 
and extending up to kiloparsec scales \citep{Hummel92}. Very Long Baseline Interferometry (VLBI) images of the source
show a complex motion pattern, with jet components moving from apparent subluminal to 
superluminal velocities \citep{Otterbein1998}.  Monitoring with MOJAVE (Monitoring Of Jets in Active galactic nuclei with VLBA Experiments)\footnote{http://www.physics.purdue.edu/astro/MOJAVE/sourcepages/0836+710.shtml} has shown apparent jet speeds up to $\beta_{app} = 25$ \cite{Lister2009}.  
Internal
 structure of the jet in 0836+710  has been investigated at 1.6 and 5 GHz using observations with the VLBI Space Observatory Programme (VSOP) \cite{Lobanov1998},  suggesting a helical structure for the jet.   A more recent study of this source with VLBI data confirmed that the helical structures observed  in the jet of 0836+710 are real and not generated artificially by the observing arrays \cite{Perucho12}. VLBA observations also suggest a spine-sheath structure for the jet of this source \cite{Asada2010}.

Blazar 0836+710  has been the subject of several X-ray studies and multiwavelength modeling to understand how X-ray emission is produced and the relationship between X-ray emission and other bands \cite{Fang2001, Foschini2006, Sambruna07, Gianni2011}.  These studies conclude that 0836+710 is a typical blazar, ultimately powered by a black hole with mass ~3$ \times$10$^{9}$ M$_{\odot}$\cite{Ghisellini10}, requiring an external source of seed photons for the high-energy Compton part of the spectrum. 

In the 1990's this blazar was detected and shown to be variable at  $>$100 MeV $\gamma$-ray energies by EGRET on board the {\it Compton Gamma Ray Observatory (CGRO)}   \cite{Thompsonetal1993}, with the name 3EG J0845+7049 in the Third EGRET Catalog \cite{Hartman99}.  Data from other  {\it CGRO} instruments (BATSE, OSSE, and COMPTEL) at lower $\gamma$-ray energies showed that the peak of the spectral energy distribution (SED) falls in the MeV energy range  \cite{Malizia2000, Collmar2006}. 
  
 A new era in $\gamma$-ray astrophysics began with the launch of the {\it Fermi Gamma-ray Space Telescope (Fermi)} in 2008 June.  This source was not bright enough in the first three months of normal mission operations
to be included in the {\it Fermi} Large Area Telescope (LAT) Bright Source List  \cite{Abdo2009}. It was, however, associated with 1FGL J0842.2+7054 
in the First LAT Catalog \citep[1FGL;][]{Abdo1FGL} and with 2FGL J0841.6+7052 in the Second LAT Catalog \citep[2FGL;][]{Nolan2012}. Regular $\gamma$-ray monitoring by {\it Fermi}-LAT showed that the source was
 not active until mid-2011.  An outburst from the source on 2011 April 3 was noted by the 
{\it Fermi}-LAT \cite{Ciprini2011}
 with an average daily flux nearly a factor of 20 greater than the average flux in the 2FGL catalog.  Its $\gamma$-ray spectrum is steep, with a photon power-law index of $2.95\pm$0.07 in 2FGL.  Even larger flares were seen in late 2011 \cite{CipriniDutka2011}, when follow-up observations with {\it Swift} were performed \cite{D'Ammando11}.   The active period for 0836+710 in $\gamma$ rays ended after 2012 January.

Motivated by the $\gamma$-ray flaring activity in 2011, we have assembled a long-term multiwavelength study of this object, with the goal of comparing its temporal and spectral properties during quiescent and flaring intervals and determining whether this high-redshift flat spectrum radio quasar (FSRQ) differs significantly from more nearby $\gamma$-ray blazars. Although this source is monitored regularly by radio telescopes and the {\it Fermi}-LAT, its coverage at other wavelengths is limited. In this paper, we present the temporal and spectral evolution of  0836+710 as observed by {\it Fermi}-LAT and contemporaneous observations at other wavelengths since the time of the {\it Fermi} launch. In Section 2, observations and data analysis are presented. Results are summarized in Section 3, and discussion is given in Section 4. Throughout the paper we use cosmological parameters  H$_{0}$ = 71~km~s$^{-1}$~Mpc$^{-1}$, $\Omega_m = 0.27$, and $\Omega_{\Lambda}=0.73$, consistent with the {\it WMAP} results \cite{Komatsu09}.

\section{Observations} 

\subsection{{\it Fermi}-LAT Analysis}

The {\it Fermi}-LAT is a $\gamma$-ray telescope operating from $20$ MeV to
$>300$ GeV. The instrument is an array of $4 \times 4$ identical
towers, each one consisting of a tracker (where the photon interacts by
pair conversion and the resulting particles are tracked) and a calorimeter (where the energy of the
pair-converted photon is measured). The entire instrument is covered
by an anticoincidence detector to help in rejecting the charged-particle
background. 
Further details on the description of LAT
are given by Atwood et al. (2009).  The LAT normally operates in sky-survey mode  in which the whole sky is observed every 2 orbits ($\sim$3 hours).

The {\it Fermi}-LAT data (100 MeV $-$ 300 GeV) considered for this analysis cover the period from 2008 August 4 (MJD 54682) to 2012 January 31 (MJD 55957). We restricted the analysis to a region of interest centered on the source and having a radius of 10$^{\circ}$.  Only
events belonging to the Source class were used. To limit contamination from the $\gamma$-ray-bright Earth limb, only events with zenith angle $<$100$^{\circ}$ were selected. The  data analysis was performed with the standard binned maximum-likelihood analysis tool {\it gtlike}, and the corresponding P7$\_$V6 Instrument Response Functions (IRF)
were used. The source model includes point sources from the 2FGL catalog \cite{Nolan2012} within 20$^{\circ}$ of 0836+710. The most important nearby source for this analysis is 2FGL~J0721.9+7120, associated with the bright, variable blazar S5 0716+714.  The background model also includes a component for the Galactic diffuse emission (gal\_2yearp7v6\_v0.fits), and an isotropic component 
(iso\_p7v6source.txt) that represents the extragalactic diffuse emission as well as residual background from cosmic-ray particles mis-identified as $\gamma$ rays.  These resources are provided separately and with the {\it Fermi}-LAT Science Tools package\footnote{Available from the {\it Fermi} Science Support Center (FSSC), http://fermi.gsfc.nasa.gov/ssc/}.

\subsubsection{Temporal Behavior}

We divided the observations into two time periods:  a quiescent period from 2008 August through 2011 February, and an active period from 2011 March through 2012 January. Light curves were extracted using likelihood analysis, fitting the source spectrum by a power-law spectral model   for these two periods.\footnote{The calculations use a Perl script (written by R. Corbet), like\_lc.pl, available on the FSSC User Contributions Web page.} The flux was evaluated by integrating the fitted
model from 100 MeV $-$ 300 GeV. A detection was defined by a Test Statistic $TS >$ 4, where $TS = 2(log L_{1} - log L_{0})$,
 where $L$ is the likelihood of the data given the model with ($L_{1}$) or without a source ($L_{0}$)  in the model \cite{Mattox96}. Although the scanning mode of {\it Fermi} produces essentially continuous observations, we have chosen to omit time intervals that have only upper limits from the plots for clarity. 
The weekly light curve  for the quiescent period, calculated with a spectral index fixed to the value in the 2FGL catalog,
is presented in  Fig. 1.  Error bars shown are statistical.  In addition,  all results have systematic uncertainties of
 10\% at 100 MeV, 5\% at 560 MeV, 10\% at 10 GeV and above \cite{Ackermann2012}.   For most time periods 0836+710 was too faint  in the LAT data for detailed studies on short time scales. 

\begin{figure*}
\centering
\includegraphics[width=155mm, angle=0]{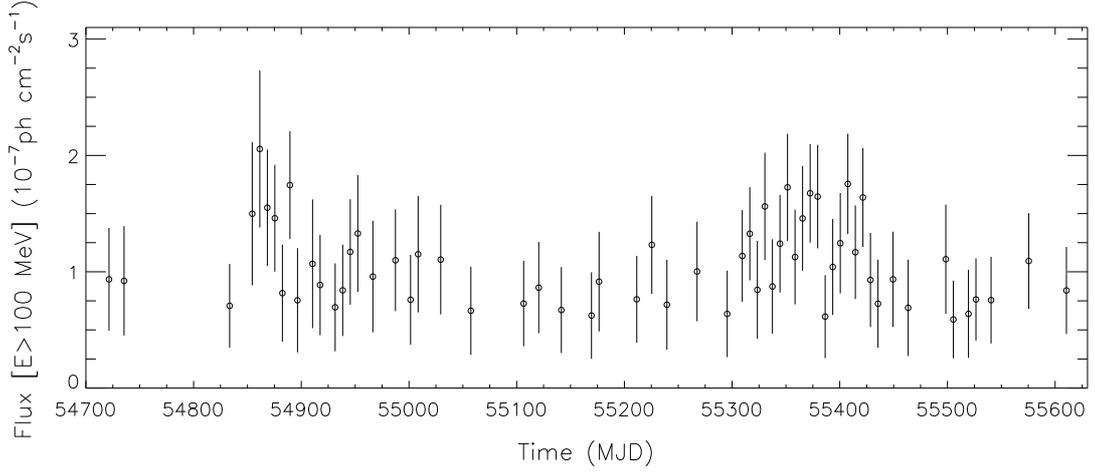}
\caption{ 
{\it Fermi}-LAT light curve for 0836+710 during the quiescent period, MJD 54682 $-$ 55621 (2008 August 4 $-$ 2011 March 1). The source flux is shown with 7-day time binning for $E >$100 MeV.  Points with $TS < $ 4 are omitted from this figure as well as
from Figs. 2 and 3.}
\end{figure*}

In Fig. 2, the weekly light curve 
for the active period is presented, again with the same fixed spectral index as in Fig. 1.
 Within this active period, the interval from 2011 October 25 (MJD 55859) to 2011 December 25 (MJD 55920) showed a significant flare, noted in the figure.  Fig. 3 shows the $E>$100 MeV flux  with one-day binning during the active period, again with the same fixed spectral index as in Fig. 1.  The flare interval shows evidence of sub-flares within the time period of enhanced emission. 
 
\begin{figure*}
\centering
\includegraphics[width=155mm, angle=0]{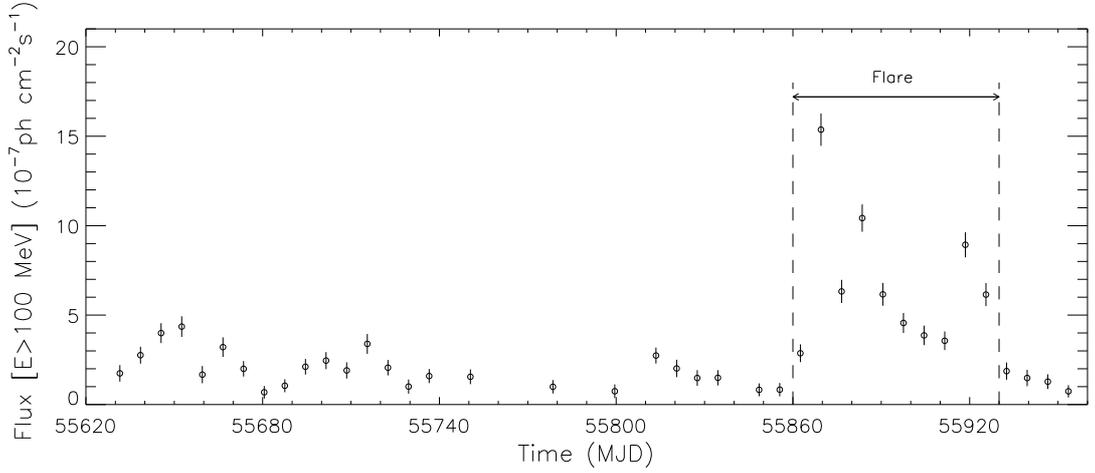}
\caption{{\it Fermi}-LAT light curve for 0836+710 during the active period, MJD 55621 $-$ 55957 (2011 March 1 $-$ 2012 January 31). The source flux is shown with 7-day time binning for $E >$100 MeV.  } 
\end{figure*}

\begin{figure*}
\centering
\includegraphics[width=155mm, angle=0]{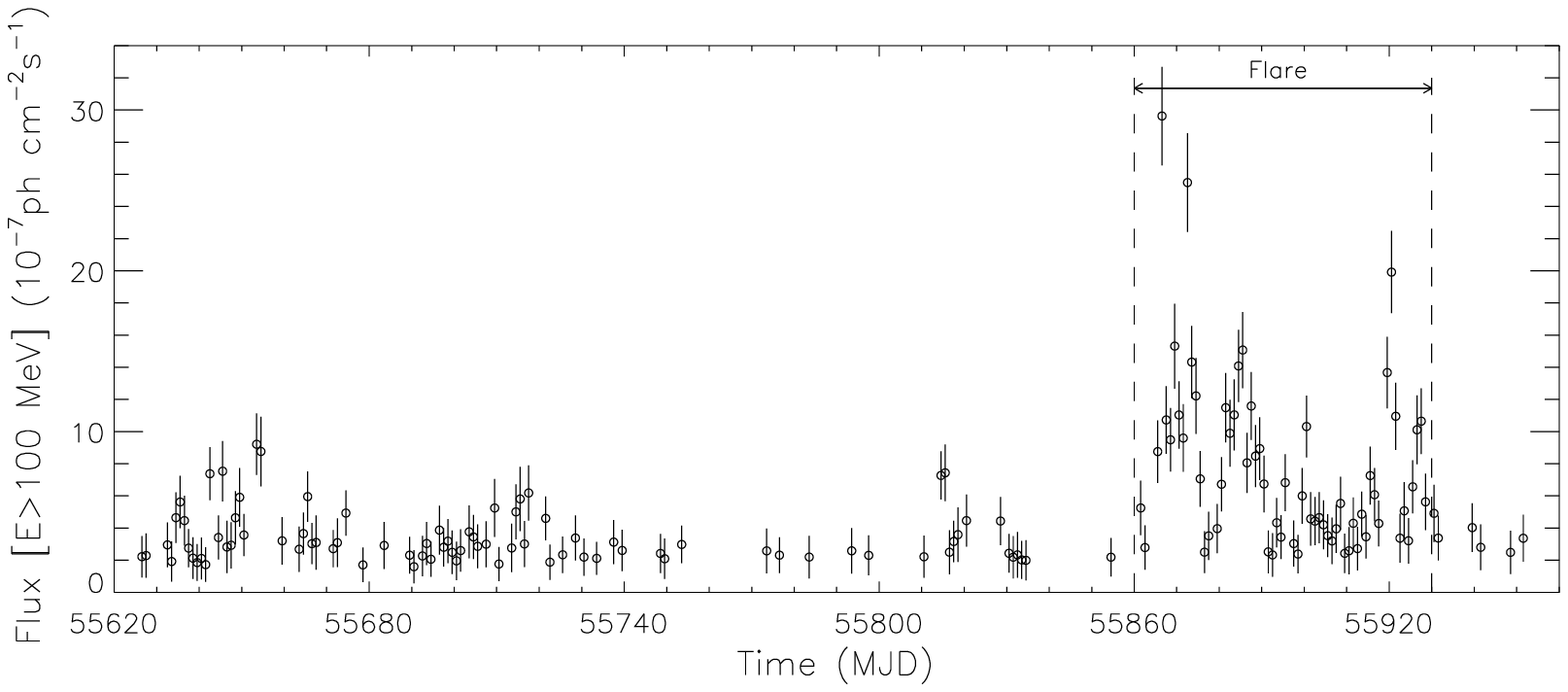}
\caption{{\it Fermi}-LAT light curve for 0836+710 during the active period, MJD 55621 $-$ 55957 (2011 March 1 $-$ 2012 January 31). The source flux is shown with 1-day time binning for $E >$100 MeV.  } 
\end{figure*}

Although its high redshift and steep energy spectrum might suggest that 0836+710 would produce too few $\gamma$ rays to enable detailed variability studies, its 2011 episode (c) flaring activity was as bright as that of many lower-redshift blazars.  In particular we note the following:

\begin{enumerate}
\item Even during the quiescent state, the blazar is regularly detected on one-week time scales, and almost always on one-month time scales (see section 3).  There are few time intervals in which the $\gamma$-ray emission is ``off.'' The particle acceleration and interaction processes that produce $\gamma$ rays must therefore be operating most of the time, not just during flaring activity. 
\item  As shown in Fig. 1, there is some indication of variability during this relatively quiet period. 
\item Like many of the blazars seen by the LAT \cite{AbdoAGNVar}, variability is seen on a variety of time scales, with much of the variation taking place on times extending over weeks or months rather than for shorter durations. 
\item In addition to the general variation, short flares are seen in 0836+710, as seen in Fig. 3. On 2011 November 1 (MJD 55866)  0836+710  flared for one day in $\gamma$ rays, reaching a flux $F$($>$100 MeV) = (3.0$\pm$0.3)$\times$10$^{-6}$ ph cm$^{-2}$s$^{-1}$, nearly 40 times brighter than the average flux in the 2FGL catalog.  The rise was a factor of 8 in a week and a factor of 4 in 24 hours or less. A similar one-day spike was seen six days later.  In order to estimate the rise times of the flare, we tried smaller time bins (1 day, 12 hours, and 6 hours) and used an exponential fit \cite{3C454flare}, producing an e-folding rise time of $\sim$0.3 days.   The counting statistics limit the accuracy of the fit.  These flares are comparable in rise time to the fastest flaring activity observed by the LAT for other blazars such as 3C 454.3 \cite{3C454flare} and PKS 1502+106 \cite{Abdo1502}.  Additional flares lasting several days appear around MJD 55885 and MJD 55920.  Because the flaring time scale sets a limit on the size of the emitting region due to light-crossing constraints, the different flaring times seen for 0836+710 suggest that the $\gamma$ rays are produced either in one small region with varying external inputs (energetic particles or seed photons) or in emitting regions of different sizes. 
\end{enumerate}

The peak daily flux for $E>$100 MeV represents an apparent isotropic luminosity of 8$\times$10$^{47}$ erg s$^{-1}$, using a luminosity distance of 18 Gpc.  This luminosity, while not the most extreme seen among $\gamma$-ray blazars, puts 0836+710 among the most powerful such objects. 

\subsubsection{Gamma-ray Spectra}

We performed the spectral analysis separately for the quiescent and active periods.  In addition to the likelihood fits for the full energy range, we computed the flux values for smaller energy intervals independently. 
This approach provides flux measurements
in each band that can be plotted along with the maximum likelihood model fit to the whole energy range,
providing a sense of how well that model describes the data. 


As a first step, we analyzed the $\gamma$-ray spectral shape of 0836+710 during the quiescent and active periods using a power-law spectral model, as this was the functional form reported in the 2FGL catalog. In the analysis for the quiescent state, 
the power-law  index $\Gamma$ for the spectrum is  (2.91$\pm$0.07) and the integral flux $F_{E>100\:\mathrm{MeV}}$ = (5.22$\pm$0.43)$\times$10$^{-8}$ ph cm$^{-2}$s$^{-1}$.  The pivot energy, for which the correlation between the photon flux and the spectral index is minimal, is 0.32 GeV.  Following the approach of the 2FGL catalog, we also tested a log-parabola spectral model\begin{equation}
\log \left (\frac{{\rm d}N}{{\rm d}E}\right ) = \log(K) - \alpha \log \left (\frac{E}{E_0} \right ) - \beta  \log^2 \left (\frac{E}{E_0} \right )
\label{eq:logparabola}
\end{equation}
The parameters are $K$ (normalization), $\alpha$ (spectral slope at $E_0$) and the curvature $\beta$, with $E_0$  an arbitrary reference energy, chosen to be the same value used in the 2FGL catalog, $E_0$ = 0.32 GeV.  In order to compare the two spectral fits, we calculated TS$_{Curv}$ as twice the difference of the log(likelihood) for the power law and the log parabola.  As shown in Fig. 4, the log-parabola form is compatible within uncertainties with a spectrum with no curvature, $\beta$ = (0.07$\pm$0.06), and provides no better fit than the power-law model, with TS$_{Curv}$=1.3. 
 
For the active time period (2011 March through 2012 January),   
the fit results in $\Gamma$ = (2.75$\pm$0.03) for a power-law model and gives a flux value $F_{E>100\: \mathrm{MeV}}$ = (2.59$\pm$ 0.09)$\times$10$^{-7}$ ph cm$^{-2}$s$^{-1}$.  Fig. 5 shows the energy spectrum of 0836+710 during this interval.  The $\gamma$-ray energy spectrum exhibited a significant deviation from a power law, as shown in the figure, with TS$_{Curv}$=43.8.  The log-parabola spectral parameters are  $\alpha$ = (2.62$\pm$0.04) and $\beta$ = (0.24$\pm$0.04).
  
  We also performed a spectral analysis for the bright  $\gamma$-ray flare (see Fig. 1 or 2) separately. 
  For this flare state,
  the power-law fit  gives $\Gamma$ = (2.65$\pm$0.04) and flux value $F_{E>100\: \mathrm{MeV}}$ = (6.05$\pm$ 0.20)$\times$10$^{-7}$ ph cm$^{-2}$s$^{-1}$.  
During this flare, the $\gamma$-ray energy spectrum also exhibited a significant deviation from a power law, as shown in Fig. 6, with TS$_{Curv}$=58.9. The log-parabola spectral parameters are $\alpha$ = (2.50$\pm$0.05) and $\beta$ = (0.32$\pm$0.05).


 The power-law representation of the spectrum in the lower-flux states suggests that the LAT energy range lies well above the energy of the peak in the SED, while one possibility for the appearance of curvature in the flaring spectrum might be that the SED peak  has shifted to higher energies (although still below the LAT energy range, making any explanation for the appearance of curvature somewhat speculative). 

\begin{figure*}
\centering
\includegraphics[width=155mm, angle=0]{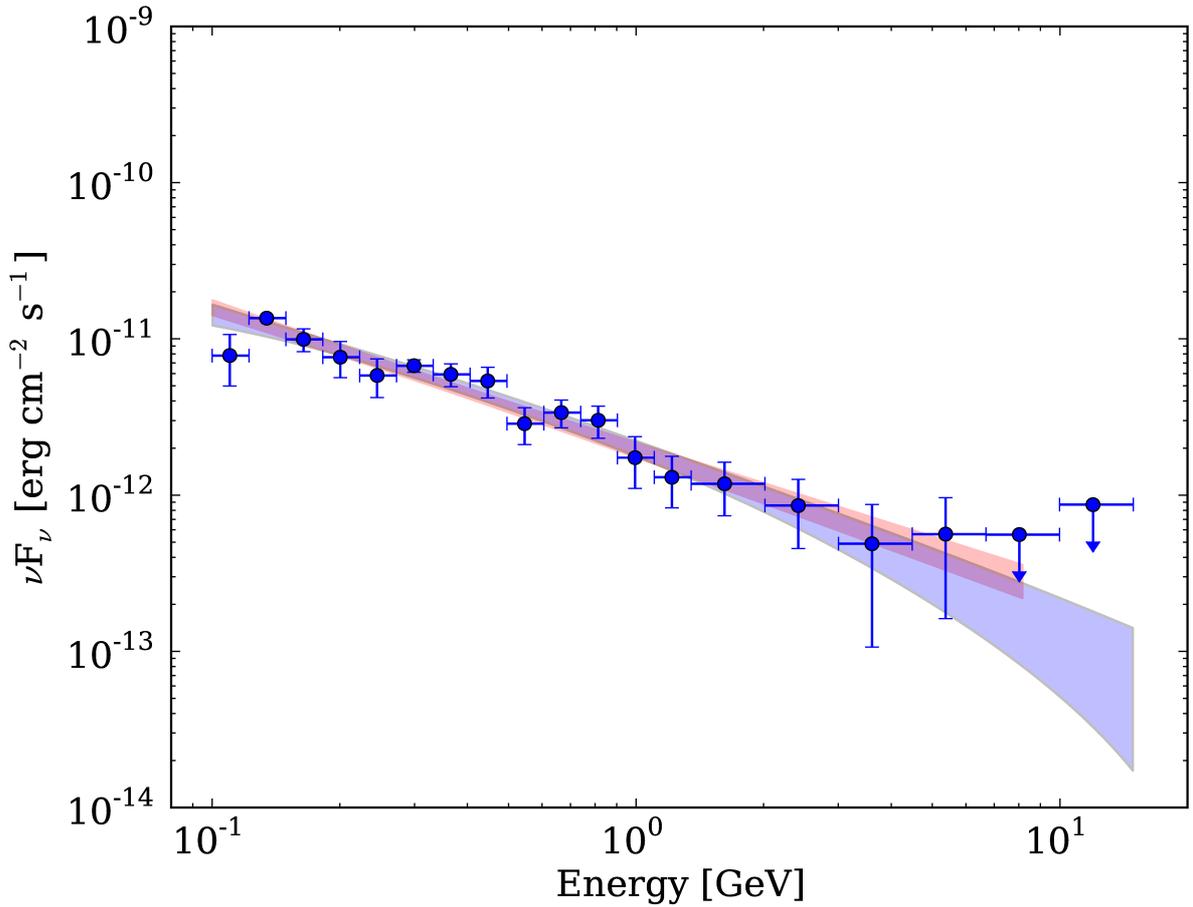}
\caption{Energy spectrum of  0836+710  during the quiescent period.  The red shaded region is the likelihood power-law fit to the entire energy range, showing the statistical uncertainty.  The blue shaded region is the likelihood log-parabola fit to the entire energy range, showing the statistical uncertainty.  The reference, or pivot, energy for each of the shaded regions is 0.32 GeV.  } 
\end{figure*}

\begin{figure*}
\centering
\includegraphics[width=155mm, angle=0]{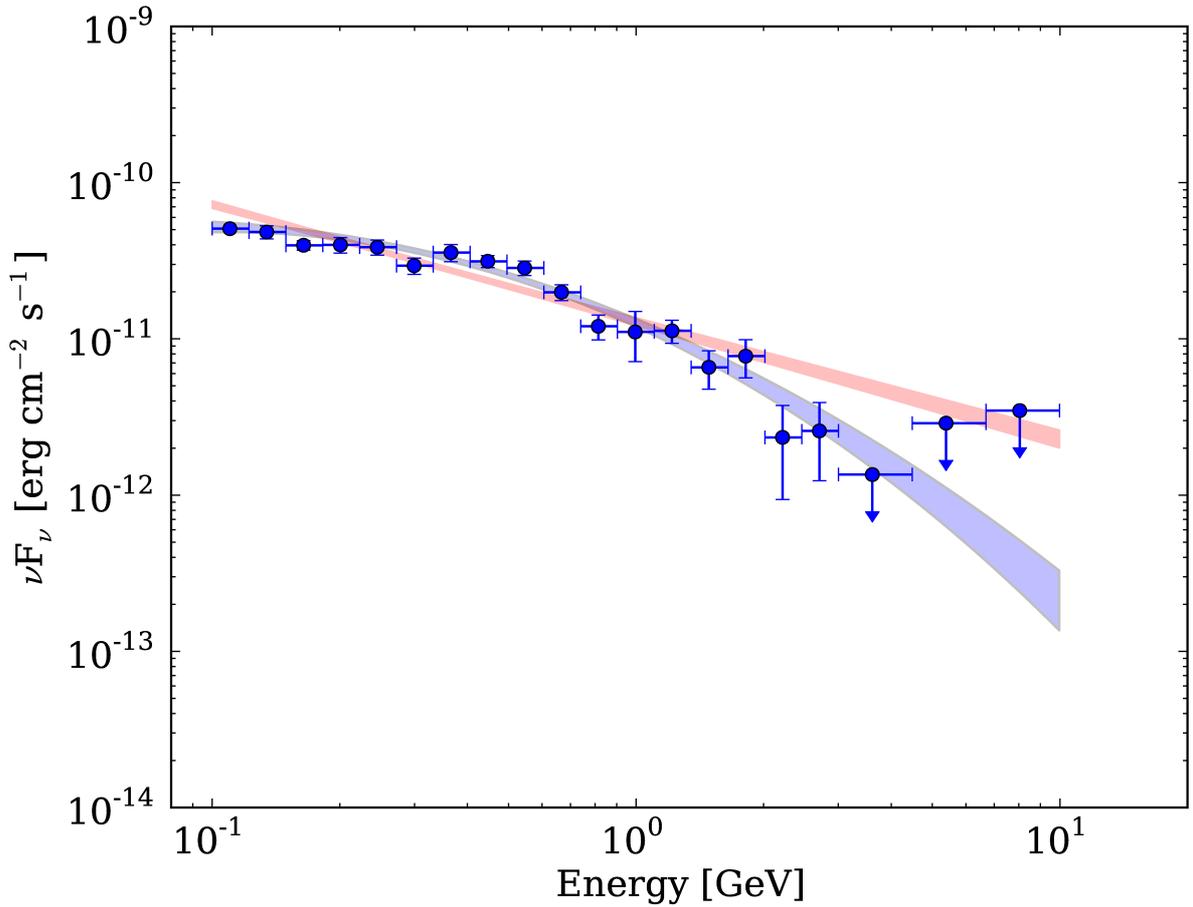}
\caption{Energy spectrum of  0836+710  during the full active period (2011 March through 2012 January). The red shaded region is the likelihood power-law fit to the entire energy range, showing the statistical uncertainty.  The blue shaded region is the likelihood log-parabola fit to the entire energy range, showing the statistical uncertainty.  The reference, or pivot, energy for each of the shaded regions is 0.32 GeV.  } 
\end{figure*}

\begin{figure*}
\centering
\includegraphics[width=155mm, angle=0]{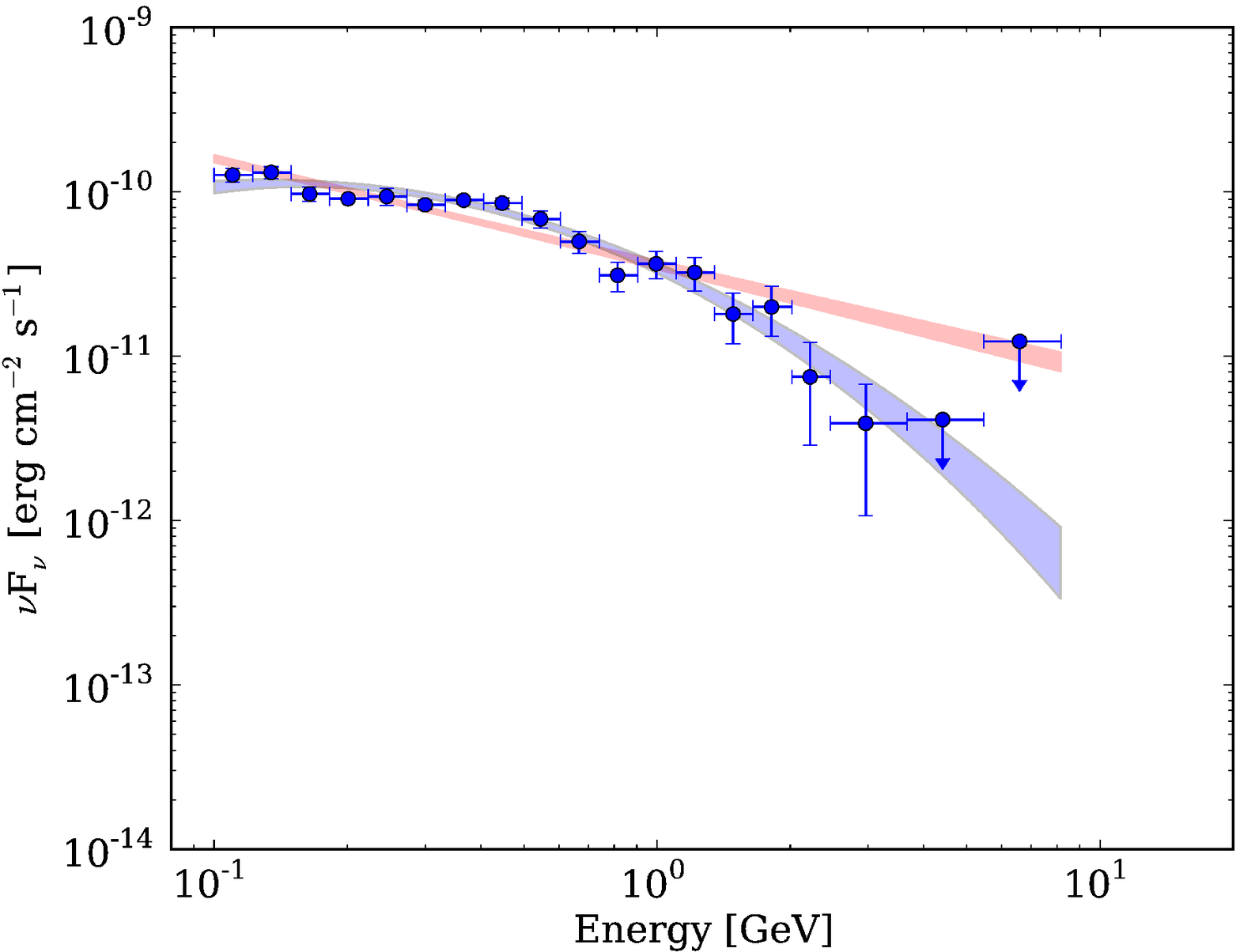}
\caption{Energy spectrum of  0836+710  during the  flare (see Fig. 1 or 2). The red shaded region is the likelihood power-law fit to the entire energy range, showing the statistical uncertainty.  The blue shaded region is the likelihood log-parabola fit to the entire energy range, showing the statistical uncertainty. The reference, or pivot, energy for each of the shaded regions is 0.32 GeV.  } 
\end{figure*}

Some {\it Fermi}-LAT blazar spectra have shown a trend toward a ``harder-when-brighter'' pattern \cite{AbdoAGNSpectra}, often with only a moderate spectral evolution, such as 3C~273 \cite{Abdo3c273} or PKS 1510$-$089 \cite{Abdo1510}, but in some cases dramatically, such as GB 1310+487 \cite{LAT1310}.  There is an indication of a ``harder-when-brighter'' pattern seen for the observations reported here, although the primary feature is the appearance of curvature in the spectrum during the active phase.  

\subsection{{\it Swift} Analysis}

\subsubsection{XRT}
The {\it Swift} X-Ray Telescope \citep[XRT;][]{Burrows05} has regularly provided X-ray coverage of blazars.  We ran the ``{\it Swift}-XRT data products generator'' \cite{Evans09}, available 
at the University of Leicester website\footnote{http://www.swift.ac.uk/user\_objects/}, to create 3 light curves, binned by observation: 
one in the 0.3$-$10 keV range and two in the 0.3$-$1.5 keV (soft) and 1.5$-$10 keV (hard) sub-ranges. 
The {\it Swift} archival data cover a long period of time, from 2006 April 1 to 
2012 April 3, and the light curves show significant variations. 
Indeed, the 0.3$-$10 keV  count rate (corrected for dead or hot pixels, the finite aperture of the extraction region, and possible pile-up) varies by a factor of 3.5 (from minimum (0.27$\pm$0.09) ct s$^{-1}$
to maximum (0.90$\pm$0.03) ct s$^{-1}$), suggesting we are observing a flaring X-ray source. This activity
is not coupled with a spectral evolution. The uniformity of the spectrum is confirmed by the fact that the hardness ratio, defined as the ratio
between the soft and the hard count rate, does not significantly vary over the course of
6 years, fluctuating around 1.1 and 1.2 (with errors of the order of 0.1).
This constant spectral behavior allows us to use all the observations simultaneously in the spectral analysis to increase the signal to noise ratio.  We used the xrtgrblc tools available within the HEASoft package released by HEASARC\footnote{http://heasarc.gsfc.nasa.gov/docs/swift/archive/grbsummary/software.html} to generate 3 sets of source and background spectral files and ancillary files. The sets correspond to the two observations in 2006, the single pointing in 2007, and all the observations afterward, respectively. Indeed, in these 3 periods the response matrix is different and the changes reflect different operating modes of the XRT. The 3 source PHA files were binned using a grouping with a minimum of 20 counts per energy bin.  This allowed us to use the $\chi^2$  statistic within XSPEC. 
We fitted the 0.3$-$10 keV spectral data with a power law and two layers of 
absorption: one fixed at the Galactic value, nH${_{Gal}}$=2.8$\times$10$^{20}$ cm$^{-2}$ \cite{Kalberla05}, and one free to vary. The additional absorption, nH${_{add}}$=(2.1$\pm$0.7)$\times$10$^{20}$ cm$^{-2}$, is 
needed. We ran the F-test within XSPEC and we found that the latter component  is significant, 
with a probability under the null hypothesis of 5$\times$10$^{-8}$. We also fitted the spectra assuming that the absorber is intrinsic 
(\emph{i.e.}, located at the source distance, $z$=2.218). In this case, the level of absorption is higher 
but with a much larger uncertainty: nH${_{int}}$=(20.2$\pm$6.0)$\times$10$^{20}$ cm$^{-2}$.  Based on the fact that 
typically blazars do not show any significant intrinsic absorption along the line of sight, we 
tend to believe that the radio estimate of the column density is simply underestimating the
amount of Galactic hydrogen column density by a factor of less than 2. Assuming that the absorption is 
Galactic, the best fit is obtained with a spectral slope of $\Gamma$=1.43$\pm$0.03. 
The observed and unabsorbed fluxes in the 0.3-10 keV range are (2.11$\pm$0.26) $\times$10$^{-11}$ erg cm$^{-2}$ s$^{-1}$ and (2.23$\pm$0.30) 
$\times$10$^{-11}$ erg cm$^{-2}$ s$^{-1}$, respectively. The conversion factor from count rate to observed flux is 4.8$\times$10$^{-11}$ erg cm$^{-2}$ ct$^{-1}$.  Results after 2007, which are contemporaneous with the observations at other wavelengths reported here, are consistent with the results accumulated over the longer time interval. 

\subsubsection{UVOT}
The  Ultraviolet/Optical Telescope \citep[UVOT;][]{Roming05} provides coverage simultaneous to the XRT for {\it Swift} observations.  Querying the {\it Swift} UVOT archive, one can notice that the source was observed with all 
the filters during each pointing only in the first years of the mission (2006 and 2007) and 
during the 2011 and 
2012 target of opportunity observations. In the period 2008-2010, only the ``filter of the day" was used. This gives us  
a partial, random coverage in the U, W1, M2, and W2 filters. We analyzed the public data using the standard tools from the {\it Swift} analysis Web page of HEASARC\footnote{http://heasarc.gsfc.nasa.gov/docs/swift/analysis/}.
For each filter, we combined all
the exposures within a single observation to estimate the  monochromatic flux (corrected for the finite aperture of the extraction region, coincidence loss, and large scale sensitivity). 
We summed both the sky images and the exposure maps using \verb+uvotimsum+. The 
photometry has been obtained running \verb+uvotsource+ and using a circular source 
extraction region (with radius varying between 3\arcsec and 5\arcsec, depending on the 
source intensity) and an annular background centered on the source (with an inner radius 
not less than 15\arcsec). 
No significant evolution is seen, although the optical and UV light curves are sparsely populated. The flux densities across the 6-year span vary
by less than 50\%. Based on the paucity of the UVOT coverage, no conclusive result can
be obtained about time variability. 
\subsubsection{BAT}
The {\it Swift} Burst Alert Telescope \citep[BAT;][]{Barthelmy05} is a wide-field hard X-ray telescope operating in the energy range from 14$-$195 keV.  The BAT results used here are from the 70-month catalog\footnote{http://heasarc.gsfc.nasa.gov/docs/swift/results/bs70mon/SWIFT\_J0841.4p7052}.  The average flux in the BAT energy range is (6.8$\pm$0.4) 
$\times$10$^{-11}$ erg cm$^{-2}$ s$^{-1}$.  The source has shown no significant variability within this time frame, which ended in 2010 September, before the start of the  $\gamma$-ray flaring activity. The power-law spectrum in the BAT catalog has photon index $\Gamma$=1.59$\pm$0.08, thus marginally steeper than the XRT spectrum. 

\subsection{Optical Observations}

R-band optical data come from the blazar monitoring program of the Abastumani Observatory \cite{Abastumani}, reduced using their standard software. The R-band photometry for 0836+710 was obtained by comparison to two nearby reference stars recommended for photometry by Villata et al. (1997)\footnote{Stars B and C in http://www.lsw.uni-heidelberg.de/projects/extragalactic/charts/0836+710.html}. 

\subsection{Radio Observations}

0836+710 is one of the blazars studied as part of the F-GAMMA (Fermi-GST AGN Multi-frequency Monitoring Alliance) radio monitoring program \cite{F-Gamma}, with an overall frequency range spanning  2.64\,GHz to 142\,GHz. The lower frequency radio data (2.6, 4.85, 8.35, 10.45, 14.6, 23, 32 and 43\,GHz) are obtained with the Effelsberg 100-m 
telescope, while the millimeter IRAM 30-m observations (86 and 142\,GHz) are closely coordinated with the more general flux monitoring conducted by IRAM.  Data from both programs are included here.  

\subsection{Other Recent Observations}

The {\it Wide-field Infrared Survey Explorer} ({\it WISE}; Wright et al. 2010) mapped the sky at wavelengths of 3.4, 4.6, 12, and 22 $\mu$m between 2010 January and August, during the quiescent phase for 0836+710.  The {\it WISE} all-sky catalog includes photometric information about 0836+710 in all four wavelength bands\footnote{http://wise2.ipac.caltech.edu/docs/release/allsky/}.

The {\it Planck} early release data included observations of 0836+710 in 2009 October and 2010 March, both also during the quiescent period. The SED for the source using these data has been included in Planck Collaboration (2011) and Giommi et al. (2012). The frequency range of {\it Planck} extends higher than that included in the F-GAMMA program. 

Optical polarization and 43 GHz VLBA observations of 0836+710 between 2011 April and December by Jorstad and Marscher (2013) as part of the Boston University blazar program\footnote{http://www.bu.edu/blazars/VLBAproject.html} have shown that the onset of the  $\gamma$-ray active period in 2011 April coincided with the emission of a new radio component with $\beta_{app} = 20\pm2$, together with a significant change of the optical polarization.  They argue that the $\gamma$-ray emission is produced more than 20 pc from the central black hole. 

\section{Multiwavelength Results}

\subsection{Long Term Multiwavelength Light Curve}

Although 0836+710 is monitored regularly by radio telescopes, its coverage at other wavelengths has been fairly sparse in recent years.  Fig. 7 summarizes the long-term flux history of the source, anchored in the top panel by the results from {\it Fermi}-LAT and in the bottom two panels by the F-GAMMA program.

\begin{figure*}
\centering
\includegraphics[width=155mm]{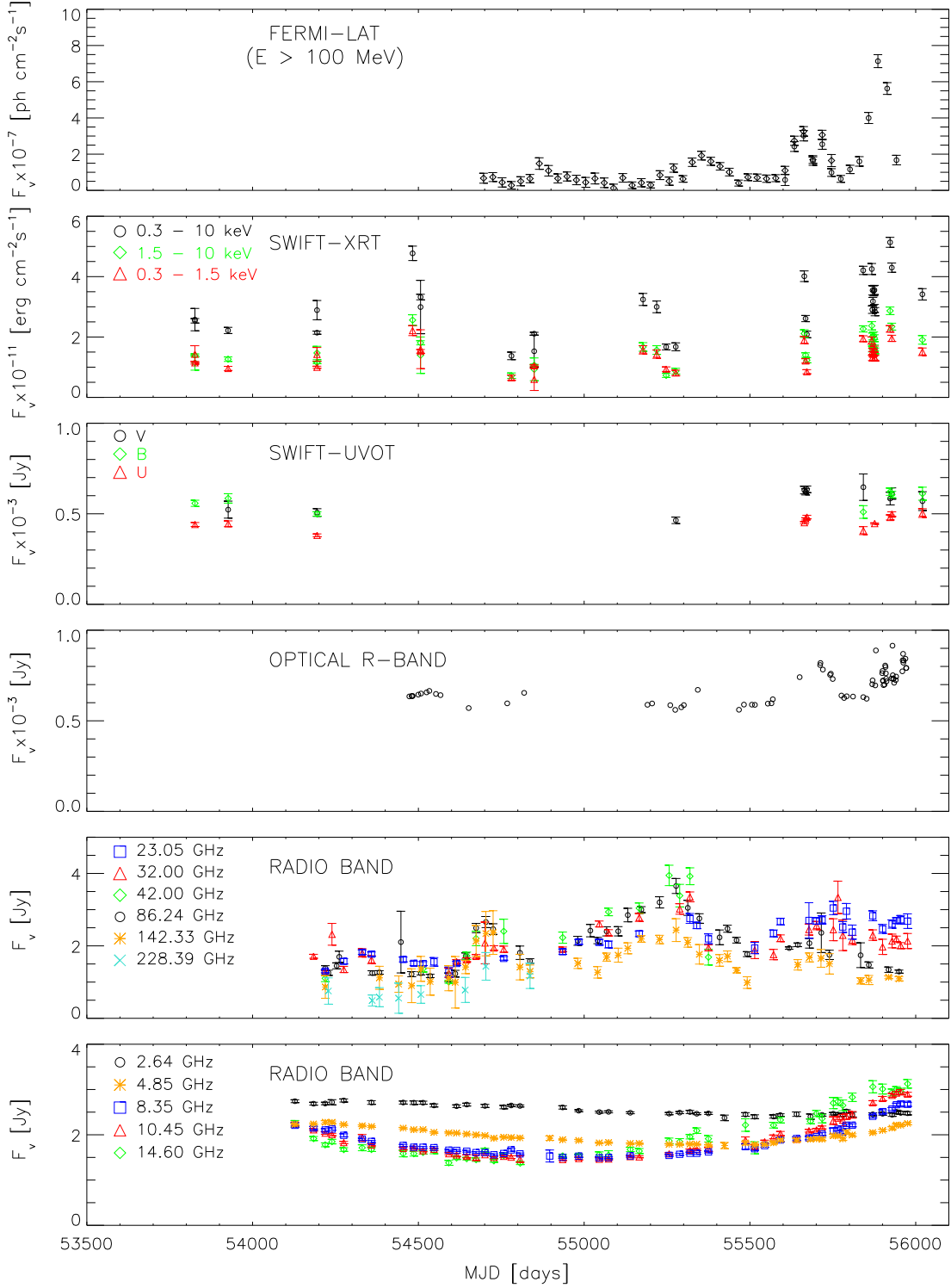}
\caption{Long term multiwavelength light curve of 0836+710.  The LAT data are accumulated in 28-day bins. See the text for descriptions of the other data sets. } 
\end{figure*}

A quantitative analysis comparing the various light curves has been carried out using the Discrete Cross Correlation Function \cite[DCCF;][]{Edelson1988}.  Although the X-ray and UV data are too sparse for any meaningful correlation analysis, the DCCF has been applied by comparing the $\gamma$-ray data with the optical and the various radio bands.

\begin{figure*}
\centering
\includegraphics[width=150mm,angle=0]{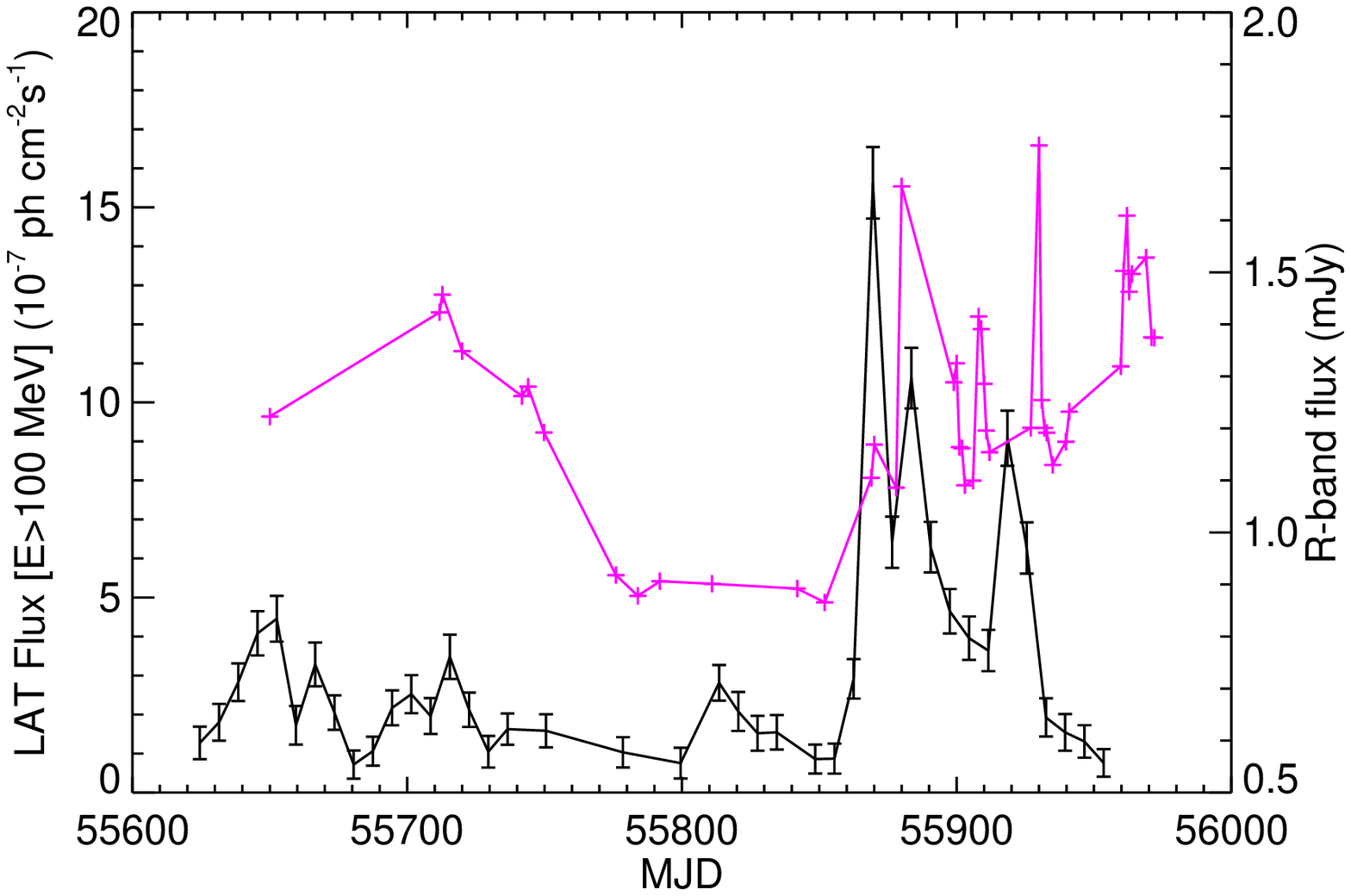}
\caption{Weekly $Fermi$-LAT $\gamma$-ray fluxes (black) compared with Abastumani R-band optical fluxes (magenta).  The data points are connected by lines to guide the eye, although only the points shown by the + symbol are actual measurements.} 
\end{figure*}

In the radio bands the source is  highly active and is flaring during the $\gamma$-ray activity, but without a simple 1 to 1 correlation with the $\gamma$ rays.  The medium/high radio frequency light curves show prominent activity throughout the whole period (starting at MJD $\sim$54500),  with flares at MJD $\sim$54700, $\sim$55250 and $\sim$55750 and with up to a factor of 4 change in amplitude.  The strongest radio flares lead the $\gamma$-ray flares by hundreds of days, making any physical connection improbable.  The lower-frequency radio bands (bottom panel of Fig. 7) do not exhibit any significant short-term flaring activity. Except at the lowest (2.64 GHz) frequency, all the lower-frequency radio bands show a generally rising trend during most of the flaring {\it Fermi} observations, a pattern that has been seen since the EGRET
era, e.g. Valtaoja \& Terasranta (1995).   Quantitatively, the radio/$\gamma$-ray DCCFs are all consistent with ``chance correlation,'' which means similar to the typical DCCF levels obtained when the 0836+710 radio light curve is correlated with the LAT light curves of 130 other AGNs for comparison \citep[see][]{Larsson}.  The radio/$\gamma$-ray situation can probably best be described as complex and ambiguous.   Statistically, active galactic nuclei seen by the LAT have correlated radio and $\gamma$-ray fluxes  \cite{Ackermann11}.  Radio/$\gamma$-ray flux comparisons for individual blazars  often show little correlation, however  \citep[e.g.][]{Abdo3C279}.  0836+710 appears to be a similar case. The appearance of high-frequency radio flares without corresponding $\gamma$-ray activity suggests that the radio emission for 0836+710 must contain one or more components not related to the source of the   $\gamma$ rays.  Such a condition could occur, for example, if a batch of synchrotron-radiating electrons lost energy before encountering a source of photons to upscatter to $\gamma$-ray energies. Another possibility is that the radio emission comes from a different region than the  $\gamma$-ray emission, at a different distance from the central black hole.

Fig. 8 shows an overlay of the Abastumani R-band optical data with the weekly $Fermi$-LAT $\gamma$-ray fluxes during the active period. There is a trend for the optical flux to be high during $\gamma$-ray enhancements, but no clear correlation of variability on shorter time scales. To evaluate the possible longer-term correlation, we constructed the DCCF using the R-band data and the 28-day $Fermi$-LAT $\gamma$-ray fluxes shown in Fig. 7. In this case, shown in Fig. 9, the optical/$\gamma$-ray DCCF indicates a correlation with time lag $\sim$ 0 $-$ 100 days ($\gamma$ rays leading optical). The 90\% confidence level in the figure is estimated independently for each lag bin and is based on the probability of 'chance correlations' in the same way as for the radio - $\gamma$-ray correlation analysis. Other FSRQs, e.g. PKS1510$-$089 \cite{Abdo1510} and 3C279 (Hayashida et al, 2012), have shown optical - $\gamma$-ray correlations with an optical emission delayed with respect to the $\gamma$ rays, strengthening the conclusion that the correlation is real, even if the value of the lag is uncertain. The correlation also provides strong support to the identification of the $\gamma$-ray source with 0836+710.

 \begin{figure*}
\centering
\includegraphics[width=150mm,angle=0]{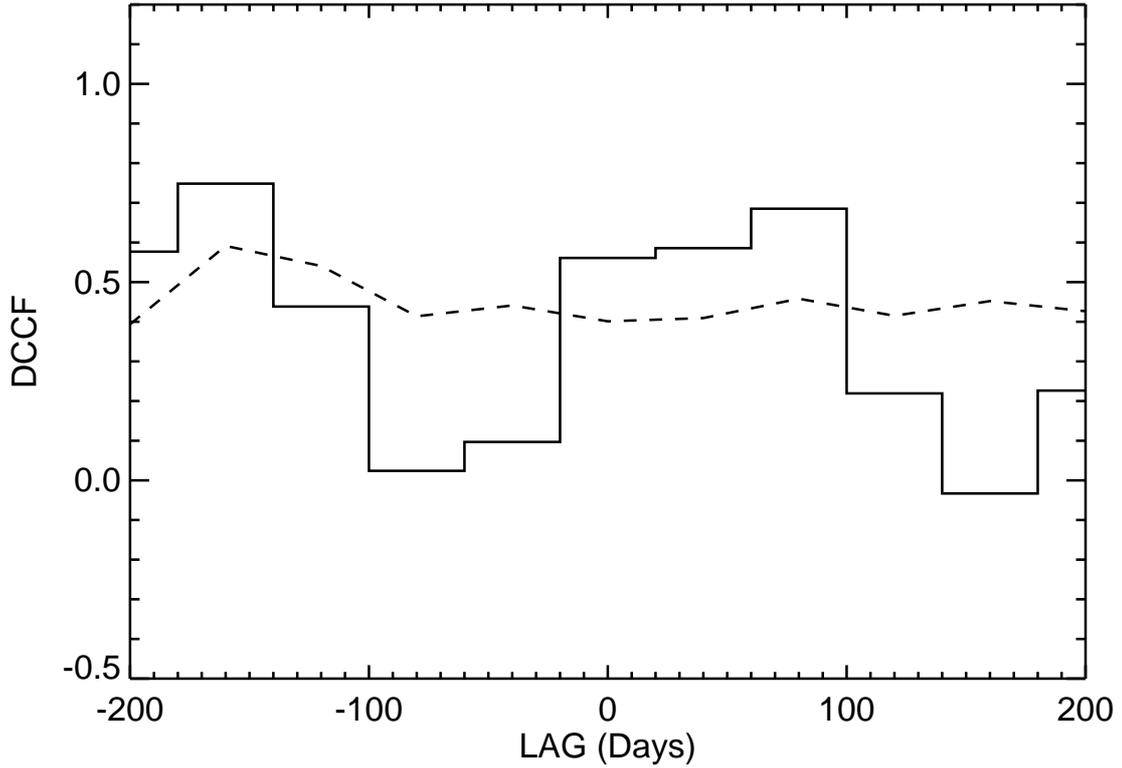}
\caption{Solid curve: Discrete Cross Correlation Function comparing the available Abastumani R-band optical data points and the 28-day $Fermi$-LAT $\gamma$-ray flux values as a function of the lag between the optical and $\gamma$-ray data.  Positive lag indicates that the $\gamma$ rays lead the optical.  Dashed curve: DCCF for a calculated correlation probability of 90\%.} 
\end{figure*}

%

\subsection{Spectral Energy Distribution}

 \begin{figure*}
\centering
\includegraphics[width=150mm,clip=true]{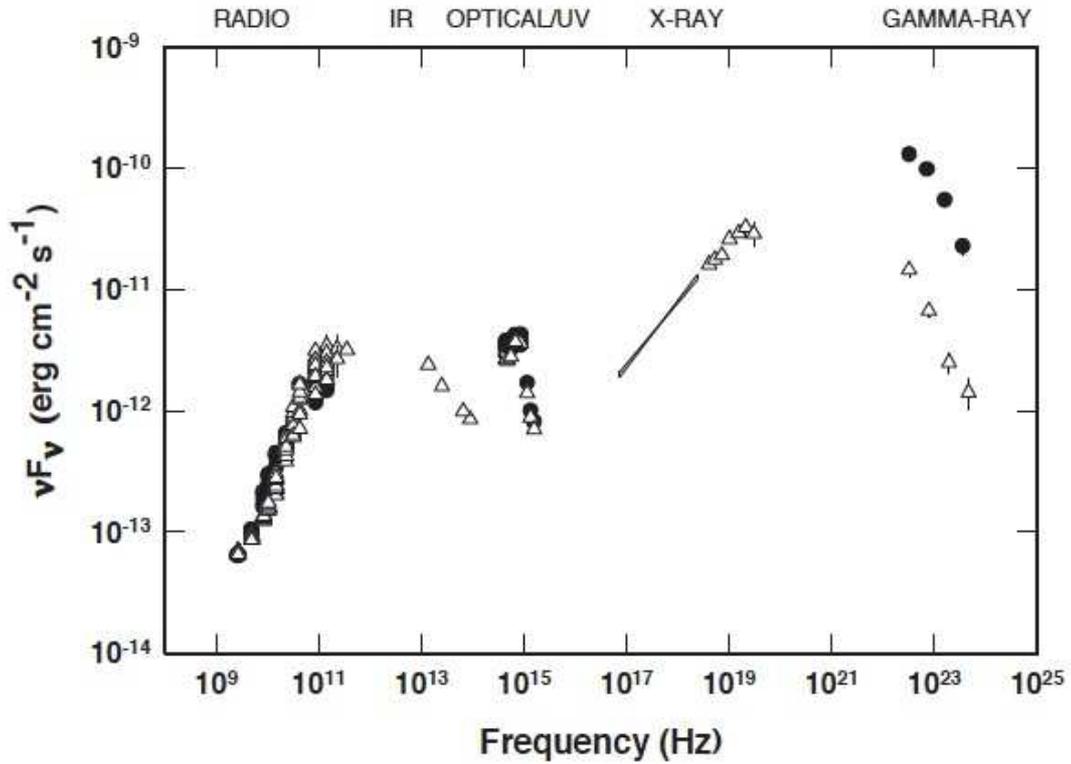}
\caption{The 0836+710  SED shows contemporaneous data for the quiescent (open symbols) and active (filled symbols) periods.  For the LAT data, the filled symbols show the spectrum during the flare, 2011 October 25 (MJD 55859) to 2011 December 25 (MJD 55920).} 
\end{figure*}

Fig. 10  shows the broadband SED of the source.  The $\gamma$-ray part of the spectrum shows the largest variability, a common occurrence for blazars.   

The combination of the {\it WISE} results with the optical and {\it Swift} UVOT observations confirms the presence of a ``blue bump'' feature, likely associated with the accretion disk, although the feature seems to have a rather narrow shape.  Such a feature has been included in SED models \cite{Sambruna07, Ghisellini10} when the blazar was in a quiescent state.  The homogeneous, one-zone Synchrotron-Self-Compton (SSC) plus External Compton (EC) model with an additional accretion disk component matches the current infrared through $\gamma$-ray data reasonably well for the quiescent state, although it fails to match the radio data, indicating that much of the radio emission does not come from the compact jet that dominates most other parts of the spectrum. Sambruna et al. (2007) suggest that a higher $\gamma$-ray flux could be explained by an injected electron spectrum that extends to higher energies than the modeled one.  Such a change would presumably push both the synchrotron and Compton peaks to higher energies, possibly resulting in a curved $\gamma$-ray spectrum during the flaring period.  The lack of MeV or far-IR observations during the flaring state makes it hard to evaluate this possibility in quantitative terms.  We have not, therefore, attempted a new SED model for this work. 

We did use the comparison of optical to $\gamma$-ray variability to obtain a clue to the primary origin of the $\gamma$-ray emission.  If the $\gamma$ rays were produced primarily by the SSC process, the $\gamma$-ray variability should go approximately as the square of the optical variability, while an EC source would predict a linear variation \cite{Maraschi}.  The complication of the accretion disk optical component makes this analysis only a rough approximation, obtained as follows:

\begin{enumerate}
\item  Extrapolating the synchrotron peak through the WISE points,  the contribution of synchrotron in the R filter is a factor of 10 lower than the observed values. 

\item The lowest flux in the R filter is 0.55 mJy; therefore the synchrotron part of this should account for 0.55/10 = $\sim$ 0.05 mJy. 

\item If we assume that the variation in the R filter (0.9 mJy at the highest peak) is solely due to synchrotron (and not to the accretion disk), than the synchrotron flux changed by a factor of (0.9$-$0.55) / 0.05 $\sim$ 7.

\item The ratio between the high and low  $\gamma$-ray flux levels is about 14, certainly smaller than would be expected from an SSC model. 

\item The variation in the synchrotron is comparable (linear) with the variation of the Compton, pointing to a model dominated by EC.
\end{enumerate}

\section{Conclusions}

This long-term multiwavelength study of 0836+710 shows no dramatic differences between this high-redshift blazar and other more nearby  $\gamma$-ray blazars.  Variability is seen on a wide range of time scales at multiple wavelengths, with the strongest variability seen in the  $\gamma$-ray band.  Although the optical and $\gamma$-ray light curves show some evidence of long-term correlation, the data do not provide a clear measurement  of the time offset. Comparing the optical and $\gamma$-ray variability shows that an EC
process is favored as the main emission mechanism of the $\gamma$-ray activity. The lack of correlation between the radio and $\gamma$-ray fluxes indicates that the radio contains components in addition to those that produce the $\gamma$-ray emission.  
The most unusual feature, the change in $\gamma$-ray  spectral shape during the flare, might be explained by a model in which the particles in the jet are accelerated to higher energies during the high-flux state.  Continued regular multiwavelength monitoring of this source will be important for resolving the ambiguities resulting from sparse coverage and for characterizing in more detail the broadband SED.

Data for all figures appearing in this paper will be posted at the $Fermi$-LAT publications Web site, http://www-glast.stanford.edu/cgi-bin/pubpub.

\begin{acknowledgements}

The \textit{Fermi} LAT Collaboration acknowledges generous ongoing support
from a number of agencies and institutes that have supported both the
development and the operation of the LAT as well as scientific data analysis.
These include the National Aeronautics and Space Administration and the
Department of Energy in the United States, the Commissariat \`a l'Energie Atomique
and the Centre National de la Recherche Scientifique / Institut National de Physique
Nucl\'eaire et de Physique des Particules in France, the Agenzia Spaziale Italiana
and the Istituto Nazionale di Fisica Nucleare in Italy, the Ministry of Education,
Culture, Sports, Science and Technology (MEXT), High Energy Accelerator Research
Organization (KEK) and Japan Aerospace Exploration Agency (JAXA) in Japan, and
the K.~A.~Wallenberg Foundation, the Swedish Research Council and the
Swedish National Space Board in Sweden.

Additional support for science analysis during the operations phase is gratefully
acknowledged from the Istituto Nazionale di Astrofisica in Italy and the Centre National d'\'Etudes Spatiales in France.

 This research is partly based on observations with the 100-m telescope of the 
MPIfR (Max-Planck-Institut f\"ur Radioastronomie) at Effelsberg. This work has 
made use of observations with the IRAM 30-m telescope. 
     
Part of this work was supported by the German
 \emph{Deut\-sche For\-schungs\-ge\-mein\-schaft, DFG\/} project
number Ts~17/2--1.
      
This work made use of data supplied by the UK {\it Swift} Science Data Centre at the University of Leicester.
      
This publication makes use of data products from the Wide-field Infrared Survey Explorer, which is a joint project of the University of California, Los Angeles, and the Jet Propulsion Laboratory/California Institute of Technology, funded by the National Aeronautics and Space Administration.

The Abastumani Observatory team acknowledges financial  support by the Georgian National Science Foundation through grant GNSF/ST09/521 4-320.
 
Part of this work was supported by the COST Action MP0905, ``Black Holes in a Violent Universe''. 

A.Akyuz acknowledges financial support by The Council of  Higher Education in Turkey. 

The authors extend special thanks to F. D'Ammando for many helpful suggestions during the preparation of this manuscript.
      
\end{acknowledgements}


\end{document}